\documentclass[epj]{svjour}
\usepackage{psfig}

\begin{document}
\title{Temperature scaling, glassiness
and stationarity in the Bak-Sneppen model}
\author{David Head
}                     % Do not remove
\institute{Department of Physics and Astronomy, JCMB King's Buildings,
University of Edinburgh, Edinburgh EH9 3JZ, United Kingdom}
\date{Received: date / Revised version: date}
\abstract{
We show that the emergence of criticality in the
locally-defined Bak-Sneppen model corresponds to separation
over a hierarchy of timescales.
Near to the critical point the model obeys scaling relations, with
exponents which we derive numerically for a one-dimensional system.
We further describe how the model can be related to the 
glass model of Bouchaud [{\em J. Phys. I France {\bf 2}, 1705 (1992)}],
and we use this insight to comment on the usual assumption of
stationarity in the Bak-Sneppen model.
Finally, we propose a general definition of self-organised criticality
which is in partial agreement with other recent definitions.
\PACS{
      {05.40.+j}{Fluctuation phenomena, random processes, and Brownian motion}   \and
      {64.60.Lx}{Self-organized criticality, avalanche effect}
     } % end of PACS codes
} %end of abstract
\maketitle

\section{Introduction}

The concept of {\em self-organised criticality} or SOC was originally
devised by Bak, Tang and Wiesenfeld as an explanation
for the apparent ubiquity of scale invariant systems in
nature~\cite{mandel1,mandel2,BTW1,BTW2,jensen,takayasu}.
Working on the assumption that such systems are `critical'
in the sense of a continuous phase transition,
Bak {\em et al.} proposed the existence of a class of
non-equilibrium models that become critical
purely under their own dynamics.
This contrasts with other equilibrium and non-equilibrium systems,
where at least one control parameter must be set to a particular
value before the critical state is attained~\cite{phase1,phase2}, which seems
unlikely in the absence of human intervention.
Despite these early claims, it soon became clear that many
SOC models {\em do} in fact require parameter tuning, but
they had been defined in such a way that the tuning
had been carried out {\em implicitly}.
For instance, the sandpile model, which is the canonical SOC system,
is usually described in a way that implicitly assumes an infinitesimal
driving rate;
for finite driving, the model is no longer critical~\cite{dickman,directed}.
The admission of implicit parameter tuning has allowed the sandpile
model to be related to more conventional non-equilibrium systems,
and for analytical techniques previously established in other
fields to be employed
(see {\em e.g.}~\cite{rs1,rs2,rs3})

Recently, attention has focused on implicit parameter tuning in a
subclass of SOC systems known as {\em extremal dynamical systems},
so-called because they are
driven at the location of the minimum (or maximum)
of some spatially varying quantity~\cite{BSRev,ip}.
It has already been observed independently by Sneppen~\cite{fint_ks}
and Vergeles~\cite{fint_mv}
that the Bak-Sneppen model, which is the simplest and perhaps best
understood extremal dynamical system~\cite{BSRev,BS,BSExpand},
implicitly assumes that a single temperature-like parameter is set
arbitrarily close to zero.
However, in our view the mechanism underlying this process was never clearly
identified, nor was the inevitability of the parameter tuning properly
stressed. 

The aims of this paper are fourfold.
Firstly, we attempt to provide as clear an explanation as
possible that implicit parameter tuning in the Bak-Sneppen model,
rather than just being possible, is in fact
{\em inevitable} if one is to have a well-defined physical system.
By `well-defined' we mean that each element evolves according to the
state of only a finite number of other elements, as opposed to the
global driving rule in the model's original invocation.
This relates to recent work on the sandpile model~\cite{dickman},
but was never properly discussed in the previous work
on this model~\cite{fint_ks,fint_mv}.
Note that throughout this paper we follow~\cite{dickman} in refering
to the setting of a parameter to zero as `fine tuning,' although
it could be argued that this is just an absolute separation of
temperature scales rather than genuine fine tuning.
Secondly, we identify the fundamental mechanism underlying this
process to be a hierarchy of timescales that diverge relative to
each other, which is similar but qualitatively different
to the separation of timescales required in other SOC models~\cite{grinstein}.
Thirdly, we show that the model can be related to a phenomenological
glass model of Bouchaud~\cite{bouchaud1},
and argue that the Bak-Sneppen model does not
reach a statistical steady state in low dimensions,
in contrary to the widely held belief
that it reaches stationarity after an `extensive transient'~\cite{BS}.
Finally, we use the insight gained from our work to propose a new way
of categorising self-organised criticality, and comment on its relationship
with other recent definitions.

This paper is arranged as follows.
In section 2 we carefully reconstruct the Bak-Sneppen model from first
principles, and show how a properly defined model with only local
interaction rules demands implicit parameter tuning,
as in the sandpile model~\cite{dickman}.
In section 3 we show that at finite temperatures the model can be
related to a simple glass model, and argue that even at its critical
point, the model does not reach a steady state.
A mean field description similar to that adopted by Bouchaud is
described and solved in section 4, and we comment on the true meaning
of self-organised criticality section 5.

\section{Reconstruction of the Bak-Sneppen model}

We begin by reconstructing the Bak-Sneppen model from first
principles, following the original arguments in~\cite{BS}.
Although some aspects of this have already been touched
upon~\cite{fint_ks,fint_mv}, the emphasis here is to demonstrate
{\em utterly unambiguously} that the implicit parameter tuning
is unavoidable, rather than just possible.

The local Bak-Sneppen model is defined as follows.
The system consists of $N$ elements, each of
which is assigned a {\em barrier}~$E_{\rm i}$\,, \mbox{$i=1\ldots N$}.
In the model's original biological context the $E_{\rm i}$ represent
{\em fitness} barriers, but they could just as
easily correspond to energy barriers on a potential energy landscape,
for example.
The values of the barriers are drawn from the time-independent
{\em prior distribution}~$\rho(E)$, which is assumed to
have no delta-function peaks so that there is a vanishing
probability of two different elements having
the same value of~$E$.
The system evolves according to two rules.
Firstly, each element becomes {\em activated} at a rate

\begin{equation}
{\rm e}^{-E_{\rm i}/T}\:\:,
\label{e:activation}
\end{equation}

\noindent{}where the constant parameter \mbox{$T>0$} has the
same units as~$E$.
An activated site $i$ is assigned a new barrier $E_{\rm i}$ drawn
from~$\rho(E)$, corresponding to a shift
to a new metastable state with a new barrier height.
Note that the activation rate for any given element is {\em independent}
of the state of the rest of the system, so
that this activation rule is strictly {\em local}.

Secondly, for every activated element, another $z$ are chosen
and also assigned new barrier values.
This interaction term mimics some form of strong coupling between
the elements, in the sense that one element changing state
drastically alters the {\em barriers} of $z$ other elements.
The way in which the $z$ interacting elements are chosen depends
upon the spatial structure of the system.
If the elements are arranged on a regular lattice,
then the most common rule is to
update the barriers of all of the nearest neighbours
of the activated element.
Hence $z$ is just the lattice coordination number.
Alternatively, the $z$ interacting elements may be chosen at
random from the remaining~$N-1$, with the connections between
the elements being randomised anew for every activation event.
In this case the system has no spatial structure and the
symbol $K$ has often been used, where \mbox{$K=z+1$}~\cite{BSMF1,BSMF2}.

The system behaviour changes qualitatively as the single
parameter \mbox{$T>0$} is varied.
These different regimes are discussed in turn below.

%% Infinitesimal T
\vspace{\baselineskip}
%\subsection{$T\rightarrow0^{+}$:}
\noindent{\em $T\rightarrow0^{+}$:}
In the limit of infinitesimal $T$,
the activation rates \mbox{${\rm e}^{-E_{\rm i}/T}$} for
different \mbox{$E_{\rm i}$}
diverge relative to each other.
That is, the element with the smallest barrier will become active on one
timescale, the one with the second smallest barrier becomes
active on another, much longer timescale, and so on.
Thus with probability one the first element to become active
will be that with the smallest barrier
(which is always unique for a
finite set of non-degenerate~$\{E_{\rm i}\}$).
This is the way in which the Bak-Sneppen model is usually
defined; indeed, this
{\em ``exponential separation of timescales''} \cite{BS}
was originally used to justify the extremal dynamics.
Although separated timescales arise in other SOC
models~\cite{grinstein},
to our knowledge this is the first time a {\em hierarchy} of timescales
has been explicitly demonstrated.

%% T small but finite
\vspace{\baselineskip}
\noindent{\em T small but finite:} The strict separation of timescales
is lost for finite $T$ and every element has a
non-vanishing probability of being the first to become active,
so the dynamics are no longer extremal.
We now demonstrate that the model is not critical in this regime.
This claim is supported by the results of numerical simulations
described below, which in turn are supported by mean field analysis.

Let $p_{i}$ denote the probability that element
$i$ becomes active before any other element in the system.
Since the time until activation follows an exponential
distribution with mean \mbox{${\rm e}^{E_{\rm i}/T}$},
it is straightforward to show that

\begin{equation}
p_{i}=\frac{{\rm e}^{-E_{\rm i}/T}}{\sum_{j=1}^{N}\,{\rm e}^{-E_{j}/T}}\:\:,
\label{e:partition}
\end{equation}

\noindent{}which obeys \mbox{$\sum_{i=1}^{N}p_{i}=1$}.
In this notation, the \mbox{$T\rightarrow0^{+}$} limit corresponds to
\mbox{$p_{i^{*}}\rightarrow1$} for the element $i^{*}$ with the smallest
barrier, and \mbox{$p_{j}\rightarrow0$} for all \mbox{$j\neq i^{*}$}.
However, $0<p_{i}<1$ for all $i$ when $T$ is finite.

The algorithm employed in the simulations was as follows.
$N$~elements were placed on a one dimensional lattice
with periodic boundary conditions.
Each element was initially assigned a barrier drawn from
the uniform prior distribution
\mbox{$\rho(E)=\{1$}~for \mbox{$0\leq E\leq1$},
0~otherwise\},
although we expect the same qualitative behaviour for other~$\rho(E)$.
For every iteration step, a single element was made active
according to the probabilities $p_{i}$ given in (\ref{e:partition}).
The active element and both of its nearest neighbours were then assigned new
barriers, so the number of interacting elements \mbox{$z=2$} here.
The timescale $t$ was normalised to $N$ activations per unit~$t$,
which differs from the usual Bak-Sneppen timescale only by
the constant factor~$N$.
Distributions were not measured until the mean barrier height
\mbox{$\langle E\rangle=\frac{1}{N}\sum_{i=1}^{N}E_{\rm i}$}
appeared to reach a steady value when plotted against
\mbox{$\log_{10}t$}.
Note that this is not a rigorous criterion for convergence and
we cannot rule out the possibility that long-range spatial correlations
may still be growing.

\begin{figure}
\centerline{\psfig{file=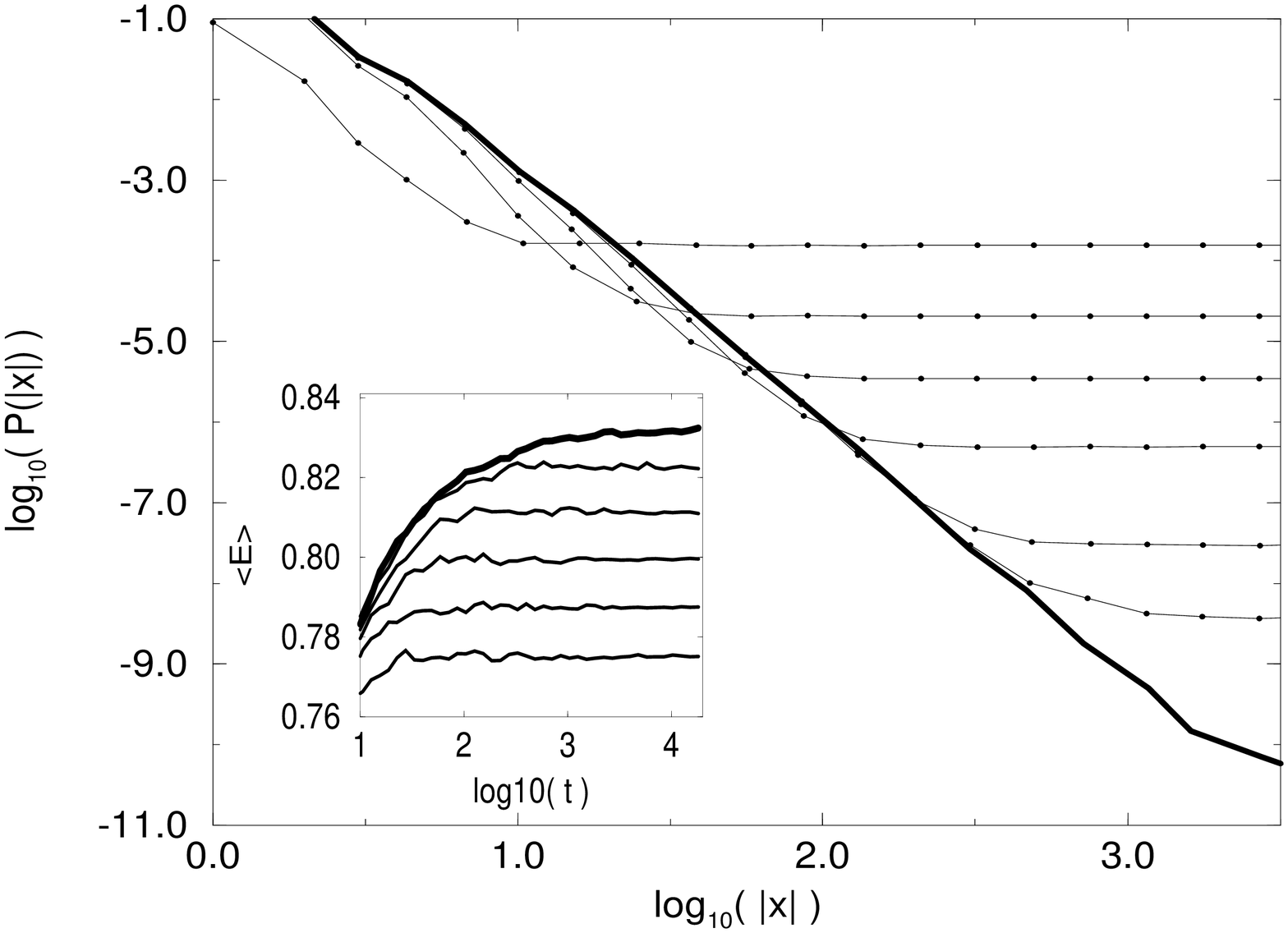,width=3.5in}}
\caption{Log-log plot of the distribution of the jump size $|x|$
between successive active sites for an \mbox{$N=10^{4}$} ring.
From bottom to top on the right hand side, the lines
refer to \mbox{$T\rightarrow0^{+}$} (thick line),
0.001, 0.002, 0.005, 0.01, 0.02 and 0.05 (thin lines), respectively.
The thick line has a slope of $-3.20\pm0.05$.
{\em (Inset)}~The mean barrier $\langle E\rangle$ against $\log_{10}t$
for $T=0^{+}$ (thick line), 0.01, 0.02, 0.03, 0.04 and 0.05 (thin lines, top
to bottom).
Each unit $t$ corresponds to $N$ activations.
Data collection commenced at \mbox{$\log_{10}t=4$}
in this and all subsequent figures.}
\label{f:jump}
\end{figure}

\begin{figure}
\centerline{\psfig{file=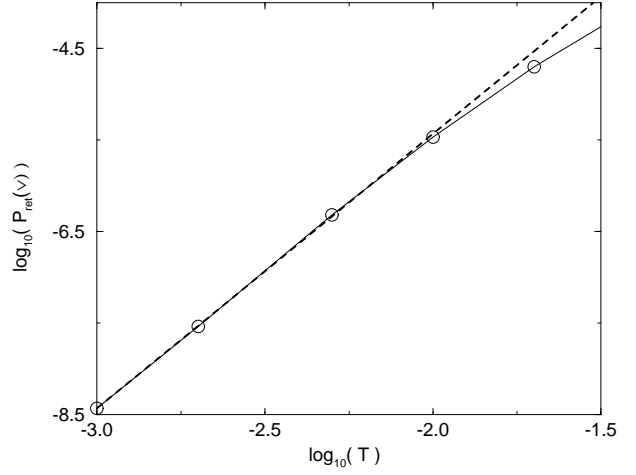,width=3.5in}}
\caption{Log-log plot of the limiting value
\mbox{$\lim_{|x|\rightarrow\infty}P_{\rm jump}(|x|)$}
as a function of $T$ for $N=10^4$.
The dashed line has a slope of~3.}
\label{f:jump_inf}
\end{figure}

To decide whether or not the system is critical for any
given value of~$T$,
we employed the usual method of extracting the spatial
and temporal correlations from the simulations
and checking to see if their tails are consistent with a power law fit.
Apart from complications arising due to finite size effects,
anything other than power law behaviour signifies a
characteristic scale and a non-critical system.
A convenient measure of spatial correlations for this
model is the number of lattice sites between two successive
active elements.
The distribution of these `jump sizes' $|x|$
for different $T$ are plotted in Fig.~\ref{f:jump}.
For \mbox{$T\rightarrow0^{+}$} we find that
\mbox{$P_{\rm jump}(|x|)\sim |x|^{-\pi}$}
with \mbox{$\pi=3.20\pm0.05$}, in accord with the known
value $3.23\pm0.02$~\cite{BSRev}.
However, $P_{\rm jump}(|x|)$ levels out at a constant $T$-dependent value
\mbox{$P_{\rm jump}(\infty)\sim A(T)$}
for finite $T$, indicating that the system is not critical.
As demonstrated in Fig.~\ref{f:jump_inf} for an 
\mbox{$N=10^{4}$} system, \mbox{$A(T)\sim T^{\alpha}$}
with \mbox{$\alpha=3.0\pm0.2$}, showing that the critical point
is indeed at \mbox{$T\rightarrow0^{+}$}.
This is consistent with the results of Sneppen~\cite{fint_ks},
but our data appears to be much smoother, allowing for a more precise
evaluation of the exponent.

\begin{figure}
\centerline{\psfig{file=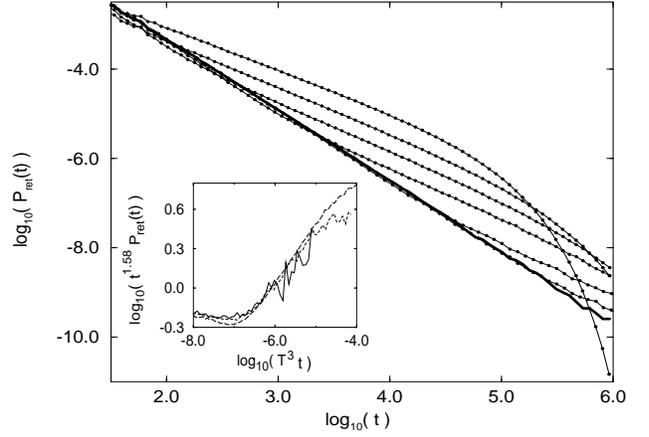,width=3.5in}}
\caption{Log-log plot of $P_{\rm ret}(t)$, the distribution of times since the
currently active site was last active, for different~$T$ on an
\mbox{$N=10^{4}$} ring.
From bottom to top in the middle of the graph, the lines refer
to \mbox{$T\rightarrow0^{+}$} (thick line),
0.001, 0.002, 0.005, 0.01, 0.02 and 0.05 (thin lines), respectively.
The thick line has a slope of $-1.58\pm0.02$.
{\em (Inset)}~Scaling plot of $t^{1.58}P_{\rm ret}(t)$ against
$T^3 t$ for $T=0.001$ (solid line), 0.002 (dotted line) and
0.005 (dashed line).}
\label{f:return}
\end{figure}

The temporal correlations are
quantified by $P_{\rm ret}(t)$, the distribution of
times $t$ since the currently active site was last active,
which is plotted in Fig.~\ref{f:return}.
Again the data for finite $T$ is clearly not power law,
and furthermore the data for small $T$ obeys a scaling function of
the form
\mbox{$P_{\rm ret}(t)\sim t^{-\alpha}\psi(T^{\beta}\,t)$}
with \mbox{$\psi(y)\rightarrow{\rm(const)}$}
as \mbox{$y\rightarrow0$}, with the exponents \mbox{$\alpha=1.58\pm0.01$}
and \mbox{$\beta=3\pm0.1$},
as demonstrated in the inset of Fig.~\ref{f:return}.

We interpret the loss of criticality for finite $T$ as follows.
For \mbox{$T\rightarrow0^{+}$}
the location of the active site jumps around the system in the
highly correlated manner characteristic of the critical state.
By contrast, when $T$ is finite there is a non-zero probability
that elements {\em arbitrarily far} from the active site will become
activated at the next time step.
Thus the active site can make large jumps to uncorrelated
regions of the system, which we relate to the loss of criticality.

%% T `Large'
\vspace{\baselineskip}
\noindent{\em $T={\mathcal{O}}(\bar{E})$:} Although there is no
longer any question of criticality far from \mbox{$T=0^{+}$}, it is
possible to relate the model to a glass model for finite~$T$.
This will be discussed fully in section~3.

%% T Infinite
\vspace{\baselineskip}
\noindent{\em $T\rightarrow\infty$:} In this limit, every element has the
same activation probability \mbox{$p_{i}=1/N$}
independent of~$E_{\rm i}$\,, so
$P(E,t)\equiv\rho(E)$ for all~$t$.

\section{Mean field analysis}

In this section we show how the results from the one dimensional
simulations are in qualitative agreement
with the solution to the mean field model.
The mean field model described here is essentially that of
Bouchaud {\em et al.} with an extra
interaction term~\cite{bouchaud2}.

Let $P(E,t){\rm d}E$ be the proportion of barriers in the range
$[E,E+{\rm d}E)$.
By adopting the usual mean field simplification of random nearest neighbours,
it is then straightforward to show that $P(E,t)$ evolves in time
according to~\cite{long}

\begin{eqnarray}
\frac{\partial P(E,t)}{\partial t}&=&
\:-\:\:\frac{{\rm e}^{-E/T}}
{\int_{0}^{\infty}{\rm e}^{-E'/T}P(E',t)\,{\rm d}E'}P(E,t)\nonumber\\
&&\:\:-z\,P(E,t)\,+\,(z\!+\!1)\rho(E)\:.
\label{e:master}
\end{eqnarray}

\noindent{}This equation can be justified by noting that
$P(E,t)$ decreases when an element changes its barrier value,
which occurs either when it becomes active,
or when it is selected as one of the $z$ interacting elements.
These two processes are described by the first and second terms on the
right hand side of (\ref{e:master}), respectively, where the prefactor to the
first term is just the continuum analogue of~(\ref{e:partition}).
Conservation of probability is ensured by the third term, which
corresponds to the $z+1$ new barriers drawn from $\rho(E)$.
We note that the treatment of Veregles is
similar~\cite{fint_mv} but with
a poorly defined timescale, resulting in factors of $N$ remaining
even {\em after} taking the continuum limit.

We have solved (\ref{e:master}) for the uniform $\rho(E)$
in the limit \mbox{$t\rightarrow\infty$}.
The expression for general $T$ is not very instructive,
but for small $T$ it simplifies to

%\begin{equation}
%P(\E,\infty)=\frac{z+1}{z}
%\left(
%1-{\rm e}^{-\E/T}
%\left\{
%\frac{\exp\left[{\frac{1}{T(z+1)}}\right]-1}
%{\exp\left[-\frac{z}{T(z+1)}\right]-1}
%\right\}
%\right)^{-1}\:\:,
%\end{equation}
%\noindent{}which for small $T$ simplifies to

\begin{equation}
P(E,\infty)\approx\frac{z+1}{z}\left(
1+
{\rm e}^{-(E-E_{\rm c})/T}
%\exp\left(-\frac{\E-\E_{\rm c}}{T}\right)
%\exp\left\{\frac{1}{T}\left[\frac{1}{z+1}-\E\right]\right\}
\right)^{-1},
\label{e:smallT}
\end{equation}

\begin{figure}
\centerline{\psfig{file=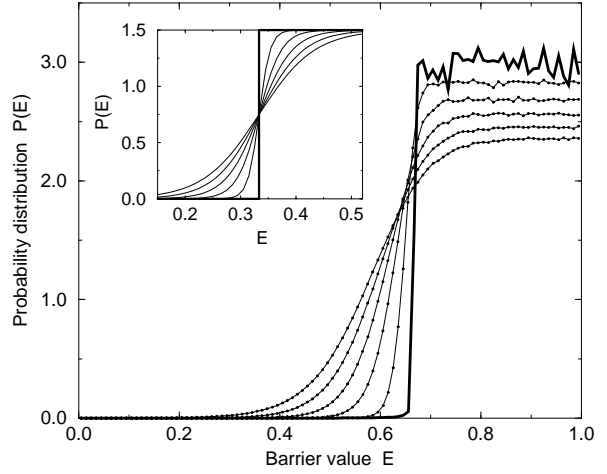,width=3.5in}}
\caption{The barrier distribution $P(E)$ for the same runs as
in Fig.~\ref{f:jump}.
From top to bottom on the right hand side, the
lines refer to \mbox{$T=0^{+}$} (thick line),
0.01, 0.02, 0.03, 0.04 and 0.05 (thin lines), respectively.
{\em (Inset)}~The corresponding mean field predictions from~(\ref{e:smallT}),
with \mbox{$z=2$}.}
\label{f:barrier}
\end{figure}

\noindent{}with \mbox{$E_{\rm c}=\frac{1}{z+1}$}\,.
As \mbox{$T\rightarrow0^{+}$} the exponential
in (\ref{e:smallT}) either blows up or decays depending on
whether $E$ is less than or greater than $E_{\rm c}$\,, respectively.
Thus $P(E,\infty)$ converges to the step function
\mbox{$\frac{z+1}{z}\,\theta(E-E_{\rm c})$},
in accord with the known solution of the mean field Bak-Sneppen
model~\cite{fint_mv,BSMF1,BSMF2,BSMF3}.
However, there is no such discontinuity for finite $T$
and $P(E,\infty)$ is smoothly varying for all \mbox{$0<E<1$},
in qualitative agreement with the simulation results
in Fig.~\ref{f:barrier} (although note that
$E_{\rm c}$ is larger in the one dimensional case).
Note that (\ref{e:master}) can also exhibit glass-like behaviour for
\mbox{$T<1$}, as fully described in~\cite{long}.

\section{Glassiness and the assumption of stationarity}

Although the model studied here
was described as the Bak-Sneppen model extended to finite temperatures,
it can equally be viewed as the simple {\em glass} model of
Bouchaud with an extra interaction
term~\cite{bouchaud1,bouchaud2,long,bouchaud3}.
The mapping to Bouchaud's model is acheived
by the two stage process of first `switching off'
the interactions, {\em i.e.} setting $z=0$, and then mapping
to a timescale $\tau$ which obeys

\begin{equation}
\frac{\partial t}{\partial\tau}
={\int_{0}^{\infty}{\rm e}^{-E'/T}P(E',t)\,{\rm d}E'} \:\:.
\end{equation}

\noindent{}The mean field equation (\ref{e:master}) then becomes

\begin{equation}
\frac{\partial P(E,\tau)}{\partial\tau}=-{\rm e}^{-E/T}P(E,\tau)
+\rho(E){\int_{0}^{\infty}{\rm e}^{-E'/T}P(E',\tau)\,{\rm d}E'}\:\:,
\end{equation}

\noindent{}which is the master equation to Bouchaud's model~\cite{bouchaud3}.
This alternative interpretation
becomes particularly relevant for values of $T$
comparable to the expected barrier height
\mbox{$\bar{E}\equiv\int\!E\rho(E)\,{\rm d}E$},
%(\,$=\frac{1}{2}$ for the uniform $\rho(\E)$),
as it is in this regime that Bouchaud's model predicts a glass transition
for a certain class of $\rho(E)$.

Since we have already described the relationship between these two models
in some detail elsewhere~\cite{long}, we will not repeat those results here.
However, it highlights what may be a very serious problem common to
almost all previous treatments of the Bak-Sneppen model.
In short, we believe there is already sufficient evidence that
the Bak-Sneppen model does {\em not} reach a statistical steady state,
just as Bouchaud's model does not reach stationarity for low temperatures.
Clearly this is contrary the the widely held assumption that it does,
so this point merits further discussion.

It is widely known from glass theory that, to truly test a system's
stationarity, it is not sufficient to measure functions of only
one time variable.
Instead one must measure a {\em two-time correlation function}
$C(t+t_{\rm w},t_{\rm w})$, which is some measure of correlation between
the state of the system at times $t_{\rm w}$ and $t_{\rm w}+t$, and show
that it does {\em not} depend on the waiting time $t_{\rm w}$
for large~$t_{\rm w}$~\cite{kob,aging}.
Within the current context, almost all of the data commonly recovered
from simulations of the Bak-Sneppen model (critical exponents,
$P(E,t)$, {\em etc.}) are all functions of at most one time variable.
For example, simply observing that $P(E,t)$ appears to approach a
limiting distribution
$P_{\infty}(E)$ does {\em not} show that the system has become stationary.
Nonetheless it appears that criteria similar to this are usually
employed to check for a statistical steady state.

To our knowledge, there has only been one instance when two-time
correlations have been actively searched for in the Bak-Sneppen model,
and this is the numerical
work of Boettcher and Paczuski in one and two dimensions~\cite{agebs}.
Remarkably, they found that the system behaviour clearly depends
on $t_{\rm w}$ for {\em all} $t_{\rm w}$ they measured.
This situation is commonly referred to as {\em aging} and is a clear indicator
of non-stationarity ({\em ie.} loss of time translational invariance).
That aging implies non-stationarity is trivial; the only question that
remains is, are Boettcher {\em et al.}'s results asymptotic, or 
is stationarity recovered at some very late time $\tau$ which is beyond
attainable simulation times? Clearly this can never be answered by
numerical simulations alone and some form of analytical treatment
would be desirable.
However, we have shown elsewhere that the
mean field model does {\em not} exhibit aging~\cite{long},
so analysis would have to limited to the difficult case of finite
dimensional systems.
This is an important issue whose resolution may help
guide attempts to find an exact solution to the model, and further work
would be desirable.

\section{Discussion and summary}

It has been suggested in~\cite{dickman} that SOC systems
correspond to absorbing state phase transitions reached in the limit of
infinitesimal driving.
Although this is almost certainly true, in our view this only explains
why SOC systems are critical, {\em not} why they are self-organised.
That is, it does not address, in sufficiently general terms,
how a model can be placed at a particular point on its phase diagram
without explicit parameter tuning.
Note that this is a separate issue as to whether or not the point also
happens to be a critical point.
An alternative definition of SOC has addressed the self-organising process
on more general terms~\cite{def1,def2},
but gives little insight into what class
of systems need to be critical to reach a statistical steady state.

In light of the work presented in this paper, we now propose 
a broad definition of SOC that is relevant to the phase diagram approach
adopted in~\cite{dickman} and throughout this paper.
We suggest that
SOC corresponds to that class of models that have a
critical point at a {\em privileged} point on their phase diagrams.
By `privileged' we mean any point at which every parameter
takes a value that has some special physical significance.
For instance, a positive definite parameter such as the rate of driving
or temperature has two privileged values, $0^{+}$ and~$\infty$, and indeed
all implicit tuning identified thus far do seem to require
infinitesimal driving or temperature~\cite{ip,grinstein}.
Similarly a conservation parameter has special points corresponding
to 100\% conservation and 100 \% dissipation.
Note that such points are {\em scale invariant} in that they do not depend
upon the chosen scale, {\em i.e.} they are absolute rather than
relative points.
If any one of these points also happens to be a critical,
then it is conceivable that the model could be placed at its critical
point `by accident' and thus be erroneously referred to as
`self-organised' critical.
It is becoming increasingly clear that this is precisely the case
is all models currently referred to as SOC.

%% Summary/conclusions
In summary, we have elucidated the mechanism behind the approach to
the critical point in the local Bak-Sneppen model.
The underlying feature is the existence of a hierarchy of timescales
that become separated as a single control parameter approaches zero.
We have used the insight gained to propose a new definition
of SOC that encompasses all cases of implicit parameter tuning observed
so far.
Furthermore we have suggested that the Bak-Sneppen model may be
non-stationary.
This is based on the relationship with a glass model for finite~$T$,
and suggests that other SOC models may also exhibit interesting behaviour
far from their critical points. We welcome study of these and related
questions.

%% ACKNOWLEDGEMENTS
\paragraph*{Acknowledgements.} The author sincerely thanks Mike Cates and
Martin Evans for useful comments and careful reading of the
manuscript, and for correspondence with Geoff Rodgers,
Kim Sneppen and Ronald Dickman.
This work was funded by EPSRC grant no. GR/M09674.

%%
%%  REFERENCES
%%

\end{document}